\documentclass[RNAAS]{aastex631}

\newcommand{\Vrot}{V_{\rm rot}}
\newcommand{\Vroti}{V_{{\rm rot}, i}}
\newcommand{\evrot}{\sigma_{V_{\rm rot}}}
\newcommand{\evroti}{\sigma_{V_{{\rm rot}, i}}}
\newcommand{\Vstar}{V_\star}
\newcommand{\Vgas}{V_{\rm gas}}
\newcommand{\Vdm}{V_{\rm DM}}
\newcommand{\mh}{M_{\rm h}}
\newcommand{\ml}{M/L}
\newcommand{\cov}{\mathsf{K}}

\begin{document}
\urlstyle{tt}

\title{Rotation curve decompositions with Gaussian Processes:\\
taking into account data correlations leads to unbiased results}

%% you desire. In v6.31 you can include a footnote in the title.

\author[0000-0001-9072-5213]{Lorenzo Posti}
\affiliation{Universit\'{e} de Strasbourg, CNRS UMR 7550, 
 Observatoire astronomique de Strasbourg, 11 rue de l’Universit\'{e}, 
 67000 Strasbourg, France}
% \email{lorenzo.posti@astro.unistra.fr}

\begin{abstract}

Correlations between velocity measurements in disk galaxy rotation curves are
usually neglected when fitting dynamical models. 
Here I show how data correlations can be taken into account in rotation curve
decompositions using Gaussian Processes. 
I find that marginalizing over correlation parameters proves critical to obtain
unbiased estimates of the luminous and dark matter distributions in galaxies.

\end{abstract}


\section{Introduction} \label{sec:intro}

Rotation curves provide robust constraints on the gravitating mass of disk galaxies,
have been historically instrumental to characterize the ``missing mass problem'', and
are useful today to link the luminous and dark masses in galaxies \citep{PFM19}.
The distribution of dark matter (DM) in galaxies is often constrained with rotation
curve decompositions employing Bayesian analyses \citep{PFM19,Li+20}. 
Usually such models assume that each point in the curve is independent, since
measurements weigh equally in standard $\chi^2$ fits that do not include correlation
terms. When extracting rotation curves from observations,
correlations between velocity measurements are typically neglected and each annulus
is treated independently. 
This, at best, is a first-order approximation, since projection effects,
resolution effects, and physical phenomena e.g. non-circular motions, can correlate
velocities in adjacent annuli.

Here, I demonstrate how to take correlations between rotation curves' datapoints
into account using Gaussian Processes (GPs) and I show that doing so is critical
to get unbiased estimates of DM halo parameters. A \texttt{python} code is made
publicly available \citep{mlpages}.

\section{Rotation curve decompositions} 

\subsection{Data}

I consider the composite H$\alpha$--HI rotation curve of the galaxy NGC2403
obtained with Fabry-Perot and radio interferometry, respectively by \cite{Daigle+06}
and \cite{Fraternali+02}.
I selected this curve from the SPARC database \citep{Lelli+16} because of its 
fine sampling and extension, so that my experiment is safe from artifacts due
poor data quality.

The rotation curve of a galaxy is
\begin{equation} \label{eq:vc}
    \Vrot^2(R) = \Vgas^2(R) + \left(\ml\right)\Vstar^2(R) + \Vdm^2(R),
\end{equation}
where $\Vrot(R)$ is the measured circular velocity, while $\Vgas(R)$, $\Vstar(R)$,
and $\Vdm(R)$ are the contributions from gas, stars, and DM.
$\Vgas$ and $\Vstar$ are determined directly from observations of the gas and stellar
surface brightness, with the unknown scaling of the mass-to-light ratio $\ml$.
$\Vdm$ is the objective of the fit and here is parametrized assuming
\citet[][hereafter NFW]{Navarro+96} profiles for the DM, i.e. with mass $\mh$ and
concentration $c$ as free parameters.
I use Eq.~(\ref{eq:vc}) to fit for the stellar mass-to-light ratio $\ml$, halo mass
$\mh$ and concentration $c$, and I take at face value the data from the SPARC catalog:
$\Vroti$ and uncertainties $\evroti$, $V_{{\rm gas},i}$, and $V_{\star,i}$ for each
radius $R_i$.

\subsection{Model without GPs: assuming independent data}

I use the Bayesian framework of \cite{PFM19} to estimate the posterior distributions of
$\theta_V=(\log\mh,\log\,c,\log\ml)$. The log-likelihood is
\begin{equation} \label{eq:chisq}
    \log\mathcal{L}(\theta_V) = -\frac{1}{2}\sum_i \frac{[\Vroti
                      -V_{\rm model}(R_i|\theta_V)]^2}{\evroti^2}.
\end{equation}
Priors are standard in this context \citep[e.g.][]{Li+20}, i.e. uninformative
for $\log\mh$, Gaussian for $\log\,c$, with mean following the $c-\mh$ relation from
cosmological simulations, and Gaussian for $\log\,(\ml)$, centered on\footnote{$L$ is
measured at $3.6\mu$m.} $\ml=0.5$.
This likelihood implicitly assumes that the datapoints are independent, because each
term in the sum depends only measurements at a specific radius $R_i$.

% I employ the ``No U-Turn Sampler'' (NUTS), implemented in \texttt{numpyro}
% \citep{numpyro1,numpyro2}, to estimate the posterior distribution.
The results are presented in Figure~\ref{fig1}, where the left-hand panel shows the
corner plot of the marginalized posterior in blue.
The posterior is well sampled and narrow, implying a tiny uncertainty i.e.
$\log\mh/M_\odot = 11.33\pm 0.02$.
The top-right panel of Fig.~\ref{fig1} shows the decomposed rotation curve where the
blue curve is the prediction of the median model. Note that I also plot a light blue
band representing the $2-\sigma$ uncertainty on the predicted $\Vrot(R)$; however,
the band is small compared to width of the blue curve and is barely visible at 
$R\gtrsim 20$ kpc.

These results are compatible with those of previous works \citep{PFM19,Li+20}.

\subsection{Model with GPs: allowing for correlated data}

To relax the assumption of independent $\Vroti$ datapoints I introduce
the covariance matrix $\cov$, whose values $\cov_{ij}$ are given by the
\emph{kernel function} $k$ as
\begin{equation} \label{eq:cov}
    \cov_{ij} = k(R_i,R_j)+\evroti^2\delta_{ij},
\end{equation}
\begin{equation} \label{eq:kernel}
    k(R_i, R_j) = A_k \exp\left[-\frac{1}{2}\left(\frac{|R_i-R_j|}{s_k}\right)^2\right]
\end{equation}
where $\delta_{ij}$ is the Kronecker delta and the parameters $A_k$ and $s_k$
are a characteristic amplitude and scale.
$k$ is maximal, implying strong correlation, for points at small distances, while it
smoothly approaches zero, implying no correlation, for points where $|R_i-R_j|\gg s_k$.
The functional form of $k(R_i, R_j)$ is arbitrarily chosen as it is common in GP studies
\citep[e.g.][]{AFM22}. Fortunately, in my testing, varying this function does not make
an impactful difference.

The new log-likelihood is
\begin{equation} \label{eq:like}
    \log\mathcal{L}(\theta_V,\theta_k) = -\frac{1}{2} {\bf V}_{\rm res}^T
                      \cov^{-1}{\bf V}_{\rm res}
                      -\frac{1}{2} \log|\mathsf{K}|,
\end{equation}
where ${\bf V}_{\rm res}$ is the vector with components
$V_{{\rm res},i}=V_{{\rm rot}, i}-V_{\rm model}(R_i|\theta_V)$, while $\cov^{-1}$ and
$|\cov|$ are the inverse and determinant of $\cov$. With this notation the problem
now becomes a GP regression, for which I use the library \texttt{tinygp} \citep{tinygp}.

\begin{figure}
\plotone{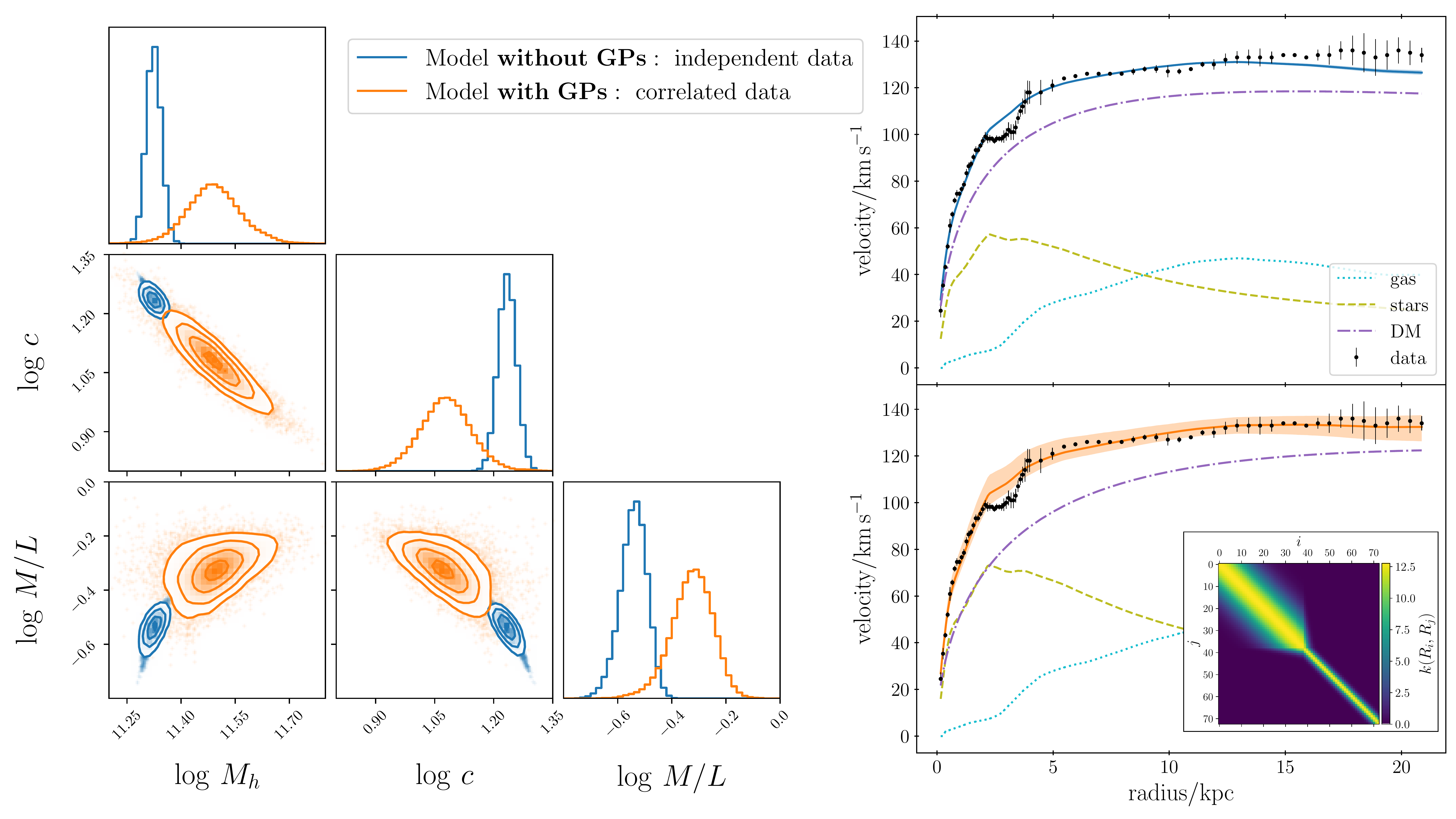}
\caption{\emph{Left-hand panel}: marginalized posterior distributions of the model without
GPs in blue, assuming independent datapoints, and of the model with GPs in orange, with
correlated datapoints. \emph{Right-hand panels}: rotation
curve decompositions of the median blue (top) and orange models (bottom) with a $2-\sigma$
uncertainty band on the predicted $\Vrot$. The inset in the bottom right panel shows
the kernel \ref{eq:kernel} of the correlation matrix.
\label{fig1}}
\end{figure}

I assume uninformative priors on the two additional free parameters $\theta_k=(\log\,A_k,
\log\,s_k)$ and I employ the same setup of the previous run.
The resulting the marginalized posterior is shown in the left-hand panel of
Fig.~\ref{fig1} as the orange model. The blue and the orange models have an incompatible mean
at $2-\sigma$, which suggests that the results of the less flexible model (blue) may be
biased.

The bottom-right panel of Fig.~\ref{fig1} shows the curve decomposition of the
median orange model. When including GPs, the model predicts larger uncertainties in
$\Vrot$ (light orange band), providing a better fit to observations, and a more massive
stellar disk, which is dominant over DM within $R\lesssim 4$ kpc, compared to the model
without GPs.
The posterior also becomes wider when including GPs, implying larger uncertainties
(e.g. $\log\mh = 11.50 \pm 0.09$), because accounting for data correlations introduces
additional degrees of freedom, which increase the model's capacity and reduce biases.
Such $\theta_k$ parameters are actually nuisance parameters that when marginalized over
yield more realistic and less biased predictions on $\theta_V$.

The inset in the bottom-right panel of Fig.~\ref{fig1} shows the resulting kernel
$k(R_i,R_j)$. The parameters $\theta_k$ are well constrained: $\log\,A_k=1.10\pm 0.26$
and $\log\,s_k/{\rm kpc}=0\pm 0.1$, i.e. points closer than $\sim 1$ kpc have
covariance of $\sim 10\,\,{\rm km^2/s^2}$ (to be compared with the typical
$\evrot^2\simeq 7 \,\,{\rm km^2/s^2}$).
The two-block shape is due to the composite nature of the curve: optical in the
center with 0.1 kpc spacing, radio in the outskirts with 0.5 kpc spacing.

\section{Conclusions}

In this paper I used state-of-the-art observations to show that assuming
independent velocity datapoints in rotation curve fits leads to biased
parameter constraints and unrealistically small uncertainties on the predicted
velocity profile.
GPs can be used to account for unknown data correlations with minimal modifications
to existing pipelines.

Deriving rotation curves from optical or radio observations is a complex task,
unlikely to result in a series of uncorrelated velocity measurements. GPs may provide
a simple and computationally cheap solution to use the wealth of literature rotation
curve data while accounting for possible underlying correlations.

\begin{acknowledgments}
I acknowledge support from the European Research Council grant agreement No. 834148.
\end{acknowledgments}

\software{tinygp \citep{tinygp},
          numpyro \citep{numpyro1},
          % arviz \citep{arviz},
          corner \citep{corner}
          }



\bibliography{biblio}{}
\bibliographystyle{aasjournal}



\end{document}